\documentclass[12pt,prd,tightenlines,showpacs,preprintnumbers,amsmath,amssymb,superscriptaddress]{revtex4}

\usepackage{natbib}
\usepackage{amssymb,amsbsy,amsmath,amsfonts}
\usepackage{graphicx}
\usepackage{epsf,epsfig,float,latexsym,amsthm,fancyhdr,rotating}
\usepackage{graphics,psfrag,longtable}
\usepackage{slashed}

\def\beq{\begin{equation}}
\def\eeq{\end{equation}}
\def\bea{\begin{eqnarray}}
\def\eea{\end{eqnarray}}
\def\beqa{\begin{equation}\begin{array}{l}}
\def\eeqa{\end{array}\end{equation}}
\def\eqlab#1{\label{Eq:#1}}

\def\Eqref#1{Eq.~(\ref{Eq:#1})}

\def\Figref#1{Fig.~\ref{Fig:#1}}

\def\quarter{\mbox{\small{$\frac{1}{4}$}}}

\def\al{\alpha}

\def\ga{\gamma} 
\def\de{\delta} \def\De{\Delta}
\def\veps{\varepsilon}

\DeclareMathOperator\arctanh{arctanh}
\begin{document}
\preprint{MITP/13-041}
\title {Chiral perturbation theory of muonic hydrogen Lamb shift: polarizability contribution}
\author{Jose Manuel Alarc\'on}
\affiliation{
Cluster of Excellence PRISMA Institut f\"ur Kernphysik, Johannes Gutenberg-Universit\"at, Mainz D-55099, Germany}
\author{Vadim Lensky}
\affiliation{
Theoretical Physics Group, School of Physics and Astronomy, 
University of Manchester, Manchester, M13 9PL, United Kingdom}
\affiliation{Institute for Theoretical and Experimental Physics,
Bol'shaya Cheremushkinskaya 25, 117218 Moscow, Russia}
\author{Vladimir Pascalutsa}
\affiliation{
Cluster of Excellence PRISMA  Institut f\"ur Kernphysik, Johannes Gutenberg-Universit\"at, Mainz D-55099, Germany}

\begin{abstract}
The proton polarizability effect in the muonic-hydrogen Lamb shift comes out as a prediction of 
baryon chiral perturbation theory at leading order
and our calculation yields for it:  $\Delta E^{(\mathrm{pol})} (2P-2S) = 8^{+3}_{-1}\, \mu$eV.
This result is consistent with most of evaluations based on dispersive sum rules, but
is about a factor of two smaller than the recent result obtained in {\em heavy-baryon} chiral perturbation theory.
We also find that the effect of $\Delta(1232)$-resonance excitation on the 
Lamb-shift is suppressed, as is the entire contribution
of the magnetic polarizability; the electric polarizability dominates.  
Our results reaffirm the point of view that the proton structure effects, beyond the charge radius, 
are  too small to resolve the `proton radius puzzle'.
\end{abstract}
\date{\today}
\maketitle

\tableofcontents

\section{Introduction}
The eight standard-deviation ($7.9\sigma$) discrepancy in the value of proton's charge radius obtained form elastic electron-proton scattering \cite{Bernauer:2010wm} and hydrogen spectroscopy \cite{Mohr:2012tt} on one hand and  from the 
muonic hydrogen ($\mu$H) spectroscopy \cite{Pohl:2010zza,Antognini:1900ns}  on the other, a.k.a.~the {\em proton charge radius
puzzle} \cite{Antognini:2013rsa,Pohl:2013yb},  is yet to meet its fully agreeable solution. 
One way to solve it is
to find an effect that would raise the $\mu H$ Lamb shift by about 310 $\mu$eV, and
 it has been suggested that proton structure could produce such an effect at $O(\alpha_{em}^5)$, e.g.~\cite{DeRujula:2010dp,Miller:2012ne}.
Most of the studies, however, derive an order of magnitude smaller effect of proton 
structure beyond the charge radius~\cite{Pachucki:1999zza,Martynenko:2005rc,Nevado:2007dd,Carlson:2011zd,Hill:2011wy,Birse:2012eb,Gorchtein:2013yga}.

The $O(\alpha_{em}^5)$ effects of proton structure in the Lamb shift are usually divided into the
effect of (i) the $3^\mathrm{rd}$ Zemach moment, (ii) finite-size recoil, and (iii) polarizabilities. The first two 
are sometimes combined into (i') the `elastic' 2$\gamma$ contribution, while the polarizability
effect is often split between  (ii') the `inelastic' 2$\gamma$ and (iii') a `subtraction' term, cf.~Table~\ref{Table:Summary}.
The `elastic' and `inelastic' 2$\gamma$ contributions are well-constrained by the available
empirical information on, respectively,  the proton form factors and unpolarized structure functions. 
The `subtraction' contribution must be modeled, and in principle one can make up a  model where the effect is
large enough to resolve the puzzle~\cite{Miller:2012ne}.

\begin{table}[b]
\centering
\begin{ruledtabular}
\begin{tabular}{ccccccccc}                          &                                                            &  Marty-                      & Nevado \&       & Carlson \&                                                  &  Birse \&          &   Gorchtein                         &   \\
                       &                Pachucki                          &       nenko                                                                                                            & Pineda             & Vanderhaeghen                                                 &  McGovern      &     {\it et al.}                                                 &         LO-B$\chi$PT            \\

($\mu$eV)                                                 &   \cite{Pachucki:1999zza}   &  \cite{Martynenko:2005rc}  & \cite{Nevado:2007dd}  & \cite{Carlson:2011zd}             & \cite{Birse:2012eb}              &  \cite{Gorchtein:2013yga}                 &   [this work]   \\
\hline
$\Delta E^{(\mathrm{subt})}_{2S}$            &  $1.8$                                                             &    $ 2.3$                              &  $--$                             &  $5.3(1.9)$                                       & $4.2(1.0)$                            &   $ -2.3 (4.6)$\footnote{adjusted value;
the original value of Ref.~\cite{Gorchtein:2013yga},  $+3.3$, is based on a different decomposition into the `elastic' and `polarizability' contributions.}                                    & $-3.0$  \\
$\Delta E^{(\mathrm{inel})}_{2S}$            &   $-13.9$                                                          &     $ -13.8$                          &   $--$                            &  $-12.7(5)$                               & $-12.7(5)$\footnote{taken from Ref.~\cite{Carlson:2011zd}.}   & $-13.0(6)$                                       & $-5.2$  \\ 
\hline
$\Delta E^{(\mathrm{pol})}_{2S}$              & $-12(2)$                                                           &    $-11.5$                            &    $-18.5$                   &  $ -7.4  (2.4)$                               & $-8.5(1.1)$                                     &  $ -15.3(5.6)$                                     & $-8.2 (^{+1.2}_{-2.5})$ \\
\end{tabular}
\caption{ Summary of available calculations
of the `subtraction' (second row), `inelastic' (third row), and 
their sum --- polarizability (last row) effects on the $2S$  level of $\mu$H. The last column represents  the $\chi$PT predictions obtained 
in this work; here the omitted effect of the $\De(1232)$-resonance
excitation is missing in the first two 
(`subtraction' and `inelastic') numbers, but it does not affect the total polarizability contribution where it is to cancel out.
\label{Table:Summary}}
\end{ruledtabular}
\end{table}

In this work we observe that chiral perturbation theory ($\chi$PT) contains definitive predictions
for all of the above mentioned $O(\alpha_{em}^5)$ proton structure effects, hence no modeling is needed,
assuming of course that $\chi$PT is an adequate theory of the low-energy nucleon structure. Some of the effects were
already assessed in the heavy-baryon variant of the theory (HB$\chi$PT), namely:
Nevado and Pineda~\cite{Nevado:2007dd}
computed the polarizability effect to leading order (LO)  [i.e., $O(p^3)]$, 
while Birse and McGovern~\cite{Birse:2012eb} computed the `subtraction' term in $O(p^4)$ HB$\chi$PT (with the caveat explained in the end of Sec.~IV).
Here, on the other hand, we work in the framework of a manifestly Lorentz-invariant variant of $\chi$PT in the baryon sector, 
referred to as B$\chi$PT~\cite{GSS89,Fuchs:2003qc,Pascalutsa:2006up,Alarcon:2012kn}. At least the LO results for nucleon polarizabilities are known
to be very different in the two variants of the theory, 
e.g., the proton magnetic polarizability is (in units of 10$^{-4}$ fm$^3$):
1.2 in HB$\chi$PT~\cite{Bernard:1995dp} vs.\ $-1.8$ in B$\chi$PT \cite{Bernard:1991rq,Lensky:2008re}. Thus, the LO effect
of the pion cloud is paramagnetic in one case and diamagnetic in the other 
(see \cite{Hall:2012iw,Lensky:2012ag} for more on HB$\chi$PT vs.\ B$\chi$PT).
Due to these qualitative and quantitative differences it is interesting to examine the
B$\chi$PT predictions for the 2$\gamma$ contributions to the Lamb shift. Here we compute the polarizability  effect 
at LO B$\chi$PT and indeed find it significantly different from 
the LO HB$\chi$PT results of Nevado and Pineda~\cite{Nevado:2007dd}, 
see~Table~\ref{Table:Summary}.

Our result for the `subtraction' and `inelastic' contributions differ from most of the previous works because we have neglected the effect of the nucleon transition into
its lowest excited state --- the $\De(1232)$. We argue however (in Sec.~III)
that the latter effect cancels out of the polarizability contribution. Thus, even though the `subtraction' and `inelastic'
values appear to be very different from the empirical values due to neglect of the $\De(1232)$ excitation, 
the polarizability contribution is not affected by this neglect.

The details of our calculation and main results are presented in the following section. Remarks on the role of the $\Delta(1232)$ excitation are given in Sec.~III. The heavy-baryon expansion
of our results is discussed in Sec.~IV. An ``effectiveness" criterion is 
applied to the HB$\chi$PT and B$\chi$PT results in Sec.~V. The conclusions are given in
 Sec.~VI. 
 Expressions for the LO $\chi$PT forward doubly-virtual proton Compton scattering (VVCS)  amplitude and pion electroproduction cross sections are given in Appendices~\ref{App:VVCS-amplitudes} and \ref{App:CrossSections}, respectively.

\section{Outline of the calculation and results}

\begin{figure}[t]
\begin{center}
\epsfig{file=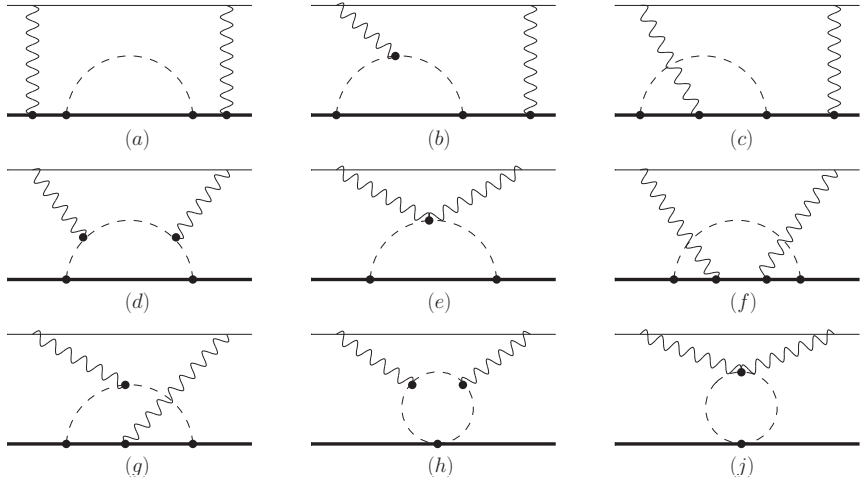,width=0.7\textwidth,angle=0}\\[1mm]
\caption{The two-photon exchange diagrams of elastic lepton-nucleon
scattering calculated in this work in the zero-energy (threshold) kinematics. Diagrams obtained from these by crossing and time-reversal
symmetry are included but not drawn. 
\label{Fig:Diagrams1}}
\end{center}
\end{figure}

We begin with the leading order chiral Lagrangian for the pion and nucleon fields, as well as the minimally-coupled photons, see e.g.~\cite{GSS89}. After a chiral rotation of the nucleon field
the Lagrangian resembles that of the chiral soliton model, see \cite{Lensky:2009uv} for details. 
As the result, the pseudovector $\pi N N$ interaction transforms into the pseudoscalar one, while a new scalar-isoscalar $\pi\pi NN$ interaction is generated.
The original and the redefined pion-nucleon Lagrangians, expanded up to the second order in the pion field, take the form:
\begin{subequations}
\bea
\mathcal{L}^{(1)}_{\pi N} & = & \overline{N}\left( i \slashed{\partial} -{M}_{N}
+ \frac{g_A}{2f_\pi} \tau^a \slashed{\partial}\,\pi^a\ga_5 - \frac{1}{4f_\pi^2} \, \tau^a \veps^{abc}  \pi^b\,\slashed{\partial} \,\pi^c
\right) N  + \mathcal{O}(\pi^3) \,,
\eqlab{expNlagran}\\
{\mathcal{L}'}_{\pi N}^{(1)} & = &  \overline{N}\left( i \slashed{\partial} -{M}_{N}
- 
i\, \frac{ g_A}{f_\pi} M_N \tau^a\pi^a\ga_5 +  \frac{g_A^2}{2f_\pi^2} M_N \pi^2
 -\frac{(g_A-1)^2}{4f_\pi^2} \, \tau^a\veps^{abc}  \pi^b\,\slashed{\partial} \,\pi^c
\right) N
 + \mathcal{O}(\pi^3)\,,\nonumber\\
\eqlab{expNlagran2}
\eea
\end{subequations}
where $N(x)$ and $M_N$ is the nucleon field and mass respectively, $\pi^a(x)$ is the pion field; $g_A\simeq 1.27$, $f_\pi\simeq 92.4$~MeV. Upon the minimal inclusion
of the electromagnetic field, the two Lagrangians give identical results
for the $O(p^3)$ Compton scattering amplitude and the isovector term proportional to $(g_A-1)^2$
does not contribute. Working with the second Lagrangian, however, simplifies a lot the evaluation of the two-loop graphs needed for the Lamb shift calculation. The resulting Feynman diagrams,
omitting crossed and time-reversed ones,
are shown in Fig.~\ref{Fig:Diagrams1}.

These graphs represent an $O(\alpha_{em}^2)$ correction to the Coulomb potential and can be treated in stationary perturbation theory.  Since the Coulomb wave function is $O(\alpha_{em}^{3/2})$,
the first-order contribution of these graphs to the energy shift is $O(\alpha_{em}^{5})$ as requested. 
As any energy transfer in the atomic system
brings in extra powers of $\alpha_{em}$, we neglect it, and hence consider strictly the zero-energy forward kinematics.
In this case the Feynman amplitude $\mathcal{M}$ in a number in momentum space, corresponding to
a potential equal to $\mathcal{M} \, \de(\vec r)$. Because of the $\de$-function only the $S$-levels are shifted:
\begin{align}\label{Eq:Energy_shift}
&\Delta E_{nS}= \phi_n^2\,  \mathcal{M},
\end{align} 
where $\phi_n^2=m_r^3 \alpha_{em}^3/(\pi n^3)$ is the hydrogen 
wave-function at the origin, for  $m_r=m_\ell\, M_p/(m_\ell+M_p)$ the reduced mass of
the lepton-proton system, and $m_\ell$, $M_p=M_N$ the corresponding  masses
of the constituents.

It is customary for the $2\gamma$ contributions to be split into leptonic and hadronic parts,
i.e.,
\begin{equation}
\mathcal{M} = \frac{e^2}{2m_\ell} \int \frac{d^4 q }{i (2\pi)^4 } \frac{1}{q^4} L_{\mu\nu}(\ell,q)\,  T^{\mu\nu}(P,q),
\end{equation}
where $e^2 = 4\pi \alpha_{em}$ is the lepton charge squared, and
\begin{equation}
L_{\mu\nu} = \frac{1}{\quarter q^4- (\ell\cdot q)^2}
\left[ q^2 \ell_\mu \ell_\nu-(q_\mu\ell_\nu +q_\nu \ell_\mu) \,\ell\cdot q 
+ g_{\mu\nu} (\ell\cdot q)^2\right]
\end{equation}
is the leptonic tensor, with $\ell$ and $q$ the 4-momenta of the lepton and the photons respectively;
$g_{\mu\nu} = \mbox{diag}(1,-1,-1,-1)$ is the Minkowski metric tensor.
The tensor $T^{\mu\nu}$ is the unpolarized VVCS amplitude,
which can be written in terms of two scalar amplitudes:
\begin{align}
&T^{\mu\nu}(P,q)= -g^{\mu \nu} \, T_1(\nu^2, Q^2) + \frac{P^\mu P^\nu }{M_p^2}  \,  T_2(\nu^2,Q^2),
\end{align} 
with $P$ the proton 4-momentum, $\nu=P\cdot q/M_p$, 
$Q^2=-q^2$, $P^2=M_p^2$. Note that the scalar amplitudes $T_{1,2}$ 
are even functions of both the photon energy $\nu$ and virtuality $Q$.
Terms proportional to $q^\mu$ or $q^\nu$ are omitted because
they vanish upon contraction with the lepton tensor.

Going back to the energy shift one obtains
\cite{Carlson:2011zd}:
\begin{align}\label{Eq:Lambshift-formula}
&\Delta E_{nS}=\frac{ \alpha_{em} \, \phi_n^2 }{4\pi^3  m_\ell} \frac{1}{i}\int\! d^3 q \int_{0}^\infty\!\!d\nu 
\frac{(Q^2-2 \nu^2)\, T_1(\nu^2,Q^2) -(Q^2+\nu^2) \, T_2(\nu^2,Q^2) }{Q^4\big[ (Q^4 /4 m_\ell^2)- \nu^2\big]}.
\end{align}

In this work we calculate the functions $T_1$ and $T_2$ by extending the B$\chi$PT calculation of 
real Compton scattering~\cite{Lensky:2009uv} to the case of virtual photons. 
We then split the amplitudes into the Born (B) and non-Born (NB) pieces:
\beq
T_i  = T_i^{(\mathrm{B})} + T_i^{(\mathrm{NB}) }.
\eeq
The Born part is defined in terms of the elastic nucleon form-factors as in, 
e.g.~\cite{Birse:2012eb,Drechsel:2002ar}:
\begin{subequations}
\bea
T_1^{(\mathrm{B})}&=&\frac{4\pi \alpha_{em}}{M_p}\left[\frac{Q^4\big(F_D(Q^2)+F_P(Q^2)\big)^2}{Q^4-4M_p^2\nu^2}-F_D^2(Q^2)\right]\,,
\label{Eq:T1Born}\\
T_2^{(\mathrm{B})}&=&\frac{16\pi \alpha_{em}M_p \,Q^2}{Q^4-4M_p^2\nu^2}\left[F_D^2(Q^2)+\frac{Q^2}{4M_p^2}F_P^2(Q^2)\right]\,.
\label{Eq:T2Born}
\eea
\end{subequations}
In our calculation the Born part was separated by subtracting the on-shell $\gamma NN$ pion loop vertex in the one-particle-reducible VVCS graphs, see diagrams (b) and (c) in Fig.~\ref{Fig:Diagrams1}.
Focusing on the $O(p^3)$ corrections (i.e., VVCS amplitude 
corresponding to the graphs in Fig.~\ref{Fig:Diagrams1}) we have explicitly verified that  the resulting NB
amplitudes  satisfy the dispersive sum rules~\cite{Bernabeu:1973zn}:
\begin{subequations}
\bea
T^{(\mathrm{NB})}_1(\nu^2,Q^2)&=& T^{(\mathrm{NB})}_1(0,Q^2) + \frac{2\nu^2}{ \pi}\int_{\nu_0}^{\infty}\!\! d\nu' \frac{\sigma_T(\nu',Q^2)}{\nu'^2-\nu^2} \label{Eq:T1disp}, \\
T^{(\mathrm{NB})}_2(\nu^2,Q^2)&=& \frac{2}{\pi}\int_{\nu_0}^{\infty}\!\! d\nu'  \frac{\nu^{\prime\, 2} Q^2}{\nu'^2+Q^2}\frac{\sigma_T(\nu',Q^2)+\sigma_L(\nu',Q^2)}{\nu'^2-\nu^2}\label{Eq:T2disp},
\eea
\eqlab{dispSRs}
\end{subequations}
with $\nu_0=m_\pi + (m_\pi^2 +Q^2)/(2M_p)$ the pion-production threshold, $m_\pi$ the pion mass, and 
$\sigma_{T(L)}$  the  tree-level cross section of pion production off the proton
induced by transverse (longitudinal) virtual photons, cf.~Appendix B.
 We hence establish that one is to calculate the `elastic' contribution 
from the Born part of the VVCS amplitudes and the `polarizability' contribution from the non-Born part, in accordance
with the procedure advocated by Birse and McGovern~\cite{Birse:2012eb}. 

Substituting the $O(p^3)$ NB amplitudes into Eq.~(\ref{Eq:Lambshift-formula}) we obtain the following
value for the polarizability correction:
\begin{align}\label{Eq:Einelexact}
\Delta E^{(\mathrm{pol})}_{2S} =-8.16~\text{$\mu$eV}.
\end{align}
This is quite different from the corresponding HB$\chi$PT
result for this effect obtained   by Nevado and Pineda~\cite{Nevado:2007dd}:
 \begin{align}\label{Eq:HBEinelexact}
\Delta E^{(\mathrm{pol})}_{2S} (\mbox{LO-HB$\chi$PT}) = -18.45 ~\text{$\mu$eV}.
\end{align}
We postpone a detailed discussion of this difference till  Sec.~\ref{Sec:HBlimit}.

\begin{figure}
\begin{center}
\epsfig{file=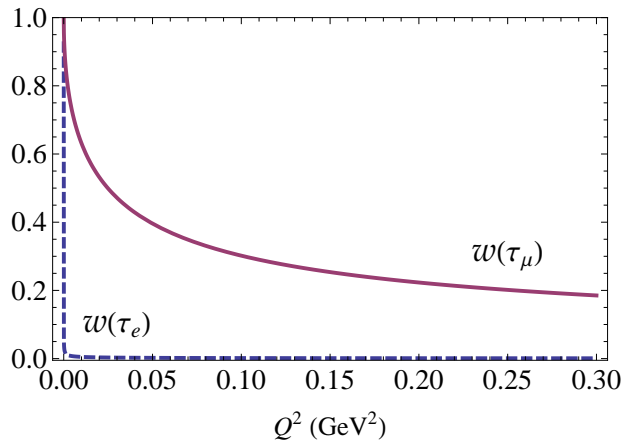,width=0.55\textwidth,angle=0}
\caption{Plot of the $Q^2$ behavior of the weighting function depending on the lepton mass. 
The blue dashed line is for the case of the electron, $w(\tau_e)$, whereas the solid purple line is for the muon, $w(\tau_\mu)$.\label{Fig:weighting}}
\end{center}
\end{figure}

It is useful to observe that
a much simpler formulae can be obtained upon making the low-energy
expansion (LEX) of the VVCS amplitude, assuming that the photon energy in the atomic system is small compared to all other scales.
To leading order in LEX, we may neglect the $\nu$ dependence in the numerator of Eq.~(\ref{Eq:Lambshift-formula}) and, 
after Wick-rotating $q$ to Euclidean hyperspherical coordinates  [i.e., setting
$
\nu = i Q \cos\chi, \quad \vec{q} = (Q \sin\chi \sin\theta\cos\varphi, \, Q \sin\chi \sin\theta\sin\varphi, Q \sin\chi \cos\theta) $] and angular integrations, 
find the following expression:
\begin{align}\label{Eq:SumRuleEstruc}
&\Delta E^{(\mathrm{pol})}_{nS}=\frac{\alpha_{em}}{\pi} \, \phi^2_n  \int_{0}^{\infty}\! \frac{dQ}{Q^2}\,  w(\tau_\ell)\,\left[ T^{(\mathrm{NB})}_1(0,Q^2)-T^{(\mathrm{NB})}_2(0,Q^2)\right],
\end{align}
with the weighting function $w(\tau_\ell)$ shown in Fig.~\ref{Fig:weighting} and given by:
\begin{align}\label{Eq:weight}
w(\tau_\ell)=\sqrt{1+\tau_\ell}-\sqrt{\tau_\ell}, \qquad \tau_\ell=\frac{Q^2}{4m_\ell^2} \,.
\end{align}
Plugging in here the LO B$\chi$PT expressions for $T_i^{(\mathrm{NB})}$
given in Appendix~\ref{App:VVCS-amplitudes}, we obtain:
\begin{align}\label{Eq:EinelLEX}
\Delta E^{(\mathrm{pol})}_{2S} =-8.20~\text{$\mu$eV},
\end{align}
 i.e., nearly the same as before the LEX, cf.~Eq.~(\ref{Eq:Einelexact}). 
 This comparison shows that the LEX is applicable in this case, i.e.:
 in the energy-shift formula of \Eqref{Lambshift-formula} 
 the $\nu$-dependence of the numerator can to an extremely 
good approximation be neglected. As shown 
in  Sec.~\ref{Sec:HBlimit}, this approximation works well
in the case of HB$\chi$PT calculation too.

To estimate the uncertainty of the LO result, we first observe that for low $Q$ the VVCS amplitudes go as:
\begin{subequations} \label{Eq:LEX}
\bea
T_1^{(\mathrm{NB})}(0,Q^2) &\simeq &  4\pi Q^2 \beta_{M1}, \\
T_2^{(\mathrm{NB})}(0,Q^2) &\simeq &  4\pi  Q^2 (\alpha_{E1}+\beta_{M1}),
\eea
\end{subequations}
where $\alpha_{E1}$ and $\beta_{M1}$ are the electric and magnetic dipole polarizabilities of the proton (hence the name ``polarizability contribution"). Given the shape of the weighting function plotted
in Fig.~\ref{Fig:weighting}, the main contribution to the integral in Eq.~(\ref{Eq:SumRuleEstruc})
comes from low $Q$'s, and therefore $\beta_{M1} $ cancels out. The dominant polarizability effect in the Lamb shift thus comes from the electric polarizability
$\alpha_{E1}$.
The B$\chi$PT physics of $\alpha_{E1}$ is such that to obtain the empirical number of about 11 (in units of 10$^{-4}$ fm$^3$), 
7 comes from LO ($\pi N$ loops) and 4 from NLO ($\pi \De$ loops), 
with uncertainty of about $\pm 1$ from the $O(p^4)$ low-energy 
constant~\cite{Lensky:2009uv}. Since in the present calculation we  
include only the LO $\pi N$ loops, we expect our value to increase in magnitude when going to the next order (i.e.,
including the $\pi \De$ loops). As the result, we replace 
the usual uncertainty of 15\% ($\simeq
 m_\pi/$GeV ) due to the higher-order effects by 
 an uncertainty of 30\% [$\simeq  (M_\De-M_p)/$GeV]  toward  the magnitude increase, anticipating in this way the effect of the $\pi \De $ loops. The 15\% uncertainty
 remains toward the magnitude decrease.
 With thus defined uncertainty, our result is: 
\begin{align}
\Delta E^{(\mathrm{pol})}_{2S}(\mbox{LO-B$\chi$PT}) =-8.2^{+1.2}_{-2.5}~\text{$\mu$eV}.
\end{align}
This is the number given in the third row of the last column 
in Table~\ref{Table:Summary}, where it can be compared to some previous results.  
Most of them agree on the polarizability contribution. As for the `inelastic'
and `subtraction' contributions, their meaningful comparison can only be made together with discussing the role of the $\De$(1232)-resonance excitation.

\section{Remarks on the $\De$(1232) contribution and `subtraction'}
Presently the most common approach to calculate the polarizability effect relies on obtaining
the VVCS amplitude from the sum rules of \Eqref{dispSRs}. Unfortunately, even a perfect
knowledge of the inclusive cross sections (or, equivalently, the unpolarized structure functions)
determines the VVCS amplitude only up to the subtraction function $T_1^{(\mathrm{NB})}(0,Q^2)$.
The total result is  therefore divided into the `inelastic' part which is determined by empirical cross sections,
and the `subtraction' terms which stands for the contribution of the subtraction function.
\begin{figure*}[tb]
\begin{center}
\epsfig{file=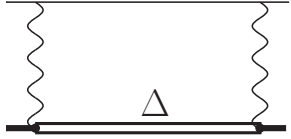,width=0.25\textwidth,angle=0}
\caption{The $\De(1232)$-excitation mechanism. Double line represents the propagator of the $\Delta$.\label{Fig:Delta}}
\end{center}
\end{figure*} 
We can also perform such a division and based on the low-energy version of the sum rules [i.e., \Eqref{SumRuleEstruc}] obtain:
\begin{subequations}
\label{Eq:decomposedLS}
\begin{align}
\Delta E^{(\mathrm{subt})}_{nS}=\frac{\alpha_{em}}{\pi} \, \phi^2_n \int_{0}^{\infty}\! \frac{dQ}{Q^2} \,  w(\tau_\ell)\,T^{(\mathrm{NB})}_1(0,Q^2) \stackrel{n=2}{=}-3.0 ~\mu\text{eV},\label{Eq:SumRuleEsubt}\\
\Delta E^{(\mathrm{inel})}_{nS}=-\frac{\alpha_{em}}{\pi} \, \phi^2_n \int_{0}^{\infty}\! \frac{dQ}{Q^2}\,  w(\tau_\ell)\,T^{(\mathrm{NB})}_2(0,Q^2) \stackrel{n=2}{=} -5.2 ~\mu\text{eV} .\label{Eq:SumRuleEinel}
\end{align}
\end{subequations}
This looks very different from the dispersive calculation, cf.~Table~\ref{Table:Summary}. The main reason for this
is the $\De(1232)$-resonance excitation mechanism shown by the graph in \Figref{Delta}. 

We have checked that the dominant, magnetic-dipole ($M1$),
part of electromagnetic nucleon-to-$\De$ transition is strongly suppressed here, as is the entire magnetic polarizability ($\beta_{M1}$) contributions, cf. discussion below Eq.~\eqref{Eq:LEX}. 
It is not suppressed in the `inelastic' and `subtraction' contributions separately, but cancels out only in the total. 
Thus, even though it is well justified to neglect the graph in \Figref{Delta} at the current level of precision, the split
into `inelastic' and `subtraction' looks unfair without it. 

In most of the dispersive calculations the cancelation of the $\De$ excitation, as well as of the entire  contribution of $\beta_{M1}$, occurs too, because the subtraction function is  at low $Q$ expressed though the empirical value for $\beta_{M1}$.
Even the HB$\chi$PT-inspired calculation of the subtraction function~\cite{Birse:2012eb}, which does not include
the $\De$(1232) explicitly, is not an exception, as a low-energy constant from $O(p^4)$ is chosen to achieve the empirical
 value for $\beta_{M1}$.  
 And even at $O(p^3)$ HB$\chi$PT, the chiral-loop contribution to $\beta_{M1}$  is --- somewhat counterintuitively --- paramagnetic 
and not too far from the empirical value, leading to a reasonable result
for the `subtraction' contribution. We take a closer look at the HB$\chi$PT prediction for the various Lamb-shift contributions in the following section. 

The central value for the `subtraction' contribution
obtained by Gorchtein~et al.~\cite{Gorchtein:2013yga} is negative,
even though the $\De$-excitation is included in their `inelastic' piece.
The quoted uncertainty of their subtraction value, however, is too large to point out any contradiction of this result with the other studies.

\section{Heavy-baryon expansion}
\label{Sec:HBlimit}

The heavy-baryon expansion, or HB$\chi$PT~\cite{JeM91a,Bernard:1995dp}, was called to salvage ``consistent power counting" which seemed
to be lost in B$\chi$PT, i.e.~the straightforward, manifestly Lorentz-invariant formulation of $\chi$PT
in the baryon sector~\cite{GSS89}. 
However, as first pointed out by Gegelia et al.~\cite{Gegelia:1999gf}, the ``power-counting violating
terms" are renormalisation scheme dependent and as such do not alter physical quantities. Furthermore,
in HB$\chi$PT they are absent only in dimensional regularisation. If a cutoff regularization is used the
terms which superficially violate power counting arise in HB$\chi$PT as well, and must be handled
in the same way as they are handled nowadays in  B$\chi$PT --- by renormalization. 

In this work for example, all such (superficially power-counting violating) terms, together with ultraviolet divergencies, are removed in the course of renormalization of the proton field, charge, anomalous magnetic moment, and mass. We use the physical
values for these parameters and hence the on-mass-shell (OMS) scheme.
This is different from the extended on-mass shell scheme (EOMS) \cite{Fuchs:2003qc}, where one starts with the parameters in the chiral limit. The physical observables, such as the Lamb shift in this case, would of course come out exactly the same in both schemes, provided
the parameters in the EOMS calculation are chosen to yield the physical 
proton mass at the physical pion mass.

Coming back to HB$\chi$PT. Despite the above-mentioned developments 
the HB$\chi$PT is still often in use. The two EFT studies of proton structure
corrections done until now \cite{Nevado:2007dd,Birse:2012eb}  are done in fact within HB$\chi$PT.
We next examine these results from the B$\chi$PT perspective. 

One of the advantages of
having worked out a B$\chi$PT result is that the one of HB$\chi$PT can easily be recovered.
We do it by expanding the expressions of Appendix~\ref{App:VVCS-amplitudes} in $\mu =
m_\pi/M_N$, while keeping the ratio of light scales $\tau_\pi = Q^2/4m_\pi^2$ 
 fixed. For the leading term
the Feynman-parameter integrations are elementary and we thus obtain
the following heavy-baryon expressions:
\begin{subequations}
\bea
T_{1}^{(\mathrm{NB})}(0,Q^2) &\stackrel{\mathrm{HB}}{=}& \frac{\al_{em} g_A^2}{4 f_\pi^2 } 
\, m_\pi  \, \left(1-\frac{1}{\sqrt{\tau_\pi} }\arctan\sqrt{\tau_\pi} \right), \\
\quad T_2^{(\mathrm{NB})}(0,Q^2) &\stackrel{\mathrm{HB}}{=}&
- \frac{\al_{em} g_A^2}{4 f_\pi^2 } 
\, m_\pi    \left( 1  -\frac{1+4\tau_\pi}{ \sqrt{\tau_\pi}} \arctan\sqrt{\tau_\pi} \right).
\eea
\end{subequations}
The first expression reproduces 
the result of Birse and McGovern (cf., $\overline{T}_1^{(3)} $ in the Appendix of~\cite{Birse:2012eb}).
We have also verified that these amplitudes correspond to the ones of Nevado and Pineda~\cite{Nevado:2007dd} at zero energy ($\nu =0$), up to a convention for an overall normalization of the amplitudes.

Substituting these expressions into \Eqref{SumRuleEstruc}, we obtain the following value
for the polarizability
contribution to the  $2S$-level shift in $\mu$H:
\begin{align}
\Delta E^{(\mathrm{pol})}_{2S}(\mbox{LO-HB$\chi$PT}) =-17.85~\text{$\mu$eV}.
\end{align}
This is slightly different from the result of Ref.~\cite{Nevado:2007dd} that
 we quote in \Eqref{HBEinelexact}, which is because of the neglected energy dependence, 
 i.e., the use of the LEX in deriving \Eqref{SumRuleEstruc} from \Eqref{Lambshift-formula}.
 Still, the difference between the exact and LEX result is well within the expected 15\% uncertainty
 of such calculation and hence we conclude that the LEX approximation works well in this case too.

Substitution to \Eqref{decomposedLS} yields the HB$\chi$PT
predictions for the `inelastic' and `subtraction' contributions:
\begin{subequations}
\label{Eq:HBdecomposedLS}
\begin{align}
\Delta E^{(\mathrm{subt})}_{2S}
(\mbox{LO-HB$\chi$PT})=1.3 ~\mu\text{eV},\\
\Delta E^{(\mathrm{inel})}_{2S}(\mbox{LO-HB$\chi$PT}) = -19.1 ~\mu\text{eV} .
\end{align}
\end{subequations}
Neglecting for a moment the difference between $\tau_\pi$ and $\tau_\mu$,  we obtain very simple
closed expressions for the Lamb shift contributions:
\begin{subequations}
\label{Eq:simpleHB}
\bea
\De E_{2S}^{(\mathrm{pol})} (\mbox{LO-HB$\chi$PT}) &\approx & \frac{\al_{em}^5 m_r^3  g_A^2}{4(4\pi f_\pi)^2} \, \frac{m_\mu}{m_\pi}
\Big( 1- 10G + 6 \ln 2\Big) = -16.1 \,\, \mu\mbox{eV}, \\
\De E_{2S}^{(\mathrm{subt})} (\mbox{LO-HB$\chi$PT}) &\approx & \frac{\al_{em}^5 m_r^3  g_A^2}{8(4\pi f_\pi)^2} 
\, \frac{m_\mu}{m_\pi}\Big( 1- 2G + 2 \ln 2\Big) = 1.1 \,\, \mu\mbox{eV},\\
\De E_{2S}^{(\mathrm{inel})} (\mbox{LO-HB$\chi$PT}) &\approx & \frac{\al_{em}^5 m_r^3  g_A^2}{8(4\pi f_\pi)^2} 
\, \frac{m_\mu}{m_\pi} \Big( 1- 18G + 10 \ln 2\Big) = -17.2 \,\, \mu\mbox{eV},
\eea
\end{subequations}
where $G\simeq 0.9160$ is the Catalan's constant. 
This should provide an impression of the parametric dependencies arising in $\chi$PT for this effect. The resulting numbers
are within the expected uncertainty for HB$\chi$PT result, and 
can in principle be easily improved in a perturbative treatment
of the pion-muon mass difference. 

So far we have been discussing the $O(p^3)$ result. 
At higher orders one in addition to the VVCS calculation needs to consider the appropriate operators from the effective lepton-nucleon Lagrangian with corresponding low-energy constants
fixed to, e.g.,  the low-energy lepton-nucleon scattering.
Birse and McGovern~\cite{Birse:2012eb} computed
the VVCS amplitude $T_1(0,Q^2)$ to order $O(p^4)$, but evaded
the consideration of the lepton-nucleon terms by introducing a ``physical cutoff" in $Q$. Hence, their resulting calculation
of the subtraction term is strongly cutoff dependent and lies, strictly speaking, 
outside the $\chi$PT framework; we refer to it as 
``HB$\chi$PT-inspired" calculation.

\section{``Effectiveness" of HB$\chi$PT vs.\  B$\chi$PT}

Although at high enough orders HB$\chi$PT and B$\chi$PT are bound
to yield the same results,  at low orders this is not necessarily so and practice
shows that especially at `predictive' orders, where
there is no free LECs to absorb the differences, HB$\chi$PT and B$\chi$PT results differ substantially, sometimes even in the sign of the total effect (cf.\ the order $p^3$ result for the magnetic polarizability of the nucleon~\cite{Lensky:2009uv,Hall:2012iw}).
The proton polarizability contribution to the Lamb shift is apparently such a case as well.
So, having found the substantial differences between  the HB$\chi$PT and  B$\chi$PT
predictions the obvious question is: which one is more reliable, if any? 

A rather common point of view is that, since HB$\chi$PT neglects only the
effects of ``higher order", any substantial disagreement only signals the importance
of higher-order effects and hence neither of the calculations should be trusted at this order.
On the other hand, it is plausible that not all the higher-order effects are large,
but only the ones
present in the B$\chi$PT calculation and dismissed in the one of HB$\chi$PT.
In support of the latter scenario is the physical principle of analyticity --- 
consequence of (micro-)causality, which
in B$\chi$PT is obeyed exactly while in HB$\chi$PT is only approximate, 
albeit improvable order by order.

Another, perhaps more quantitative criterion is the one put forward 
by Strikman and Weiss~\cite{Strikman:2009bd}. In the interpretation
of Ref.~\cite{Hall:2012iw}, it  
requires that the high-momentum contribution of finite (renormalized) loop integrals over quantities
 which are invariant under redefinitions of hadron fields should not exceed the expected uncertainty 
 of the given-order calculation. In other words, 
 the contribution from beyond the scales at which the effective theory is applicable should not exceed a natural estimate of missing higher-order effects. 
 
\begin{figure}
\begin{center}
\epsfig{file=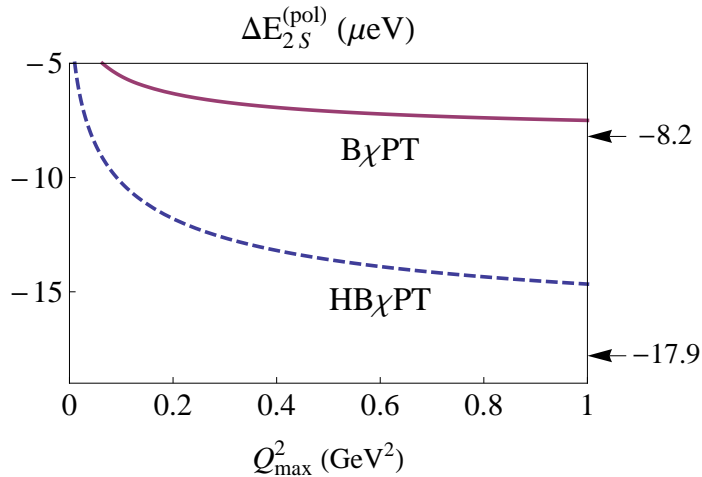,width=0.6\textwidth,angle=0}
\caption{The polarizability effect on the $2S$-level shift in $\mu$H computed
in HB$\chi$PT and  B$\chi$PT as a function of the ultraviolet cutoff $Q_{max}$. The arrows
on the right indicate the asymptotic ($Q_{max}\to\infty $) values. }
\label{Fig:Epol}
\end{center}
\end{figure}

 In our case the VVCS amplitudes are such quantities
 invariant under redefinitions of pion and nucleon fields and hence it makes sense to examine 
 \Figref{Epol}, where the polarizability effect is plotted as function of an ultraviolet cutoff $Q_{max}$
 imposed on the momentum integration in  \Eqref{SumRuleEstruc}.
The figure clearly shows that the relative size of the high-momentum contribution in
the HB$\chi$PT case is substantially larger than in B$\chi$PT. 

Assuming the breakdown scale for $\chi$PT is of order of the $\rho$-meson mass, $m_\rho = 777$~MeV, we can make a more quantitative statement. 
In the present HB$\chi$PT calculation 
the contribution from $Q> m_\rho $ is least 25\% of the total result, hence exceeding the natural expectation of uncertainty of such calculations.  
 In the B$\chi$PT case, the contribution from
 momenta above $m_\rho$ is less than 15\%,  well within the expected uncertainty.

\section{Conclusion and outlook}
Is the proton polarizability effect different in muonic versus electronic hydrogen such as to affect the charge
radius extraction? The answer is `yes'.
From the LEX formula  in Eq.~(\ref{Eq:SumRuleEstruc}), one sees that the polarizability contribution
not only affects the charge radius extraction from the Lamb shift but also that this effect is about $m_\mu/m_e\approx 200$ 
times stronger in $\mu$H than in $e$H. Indeed, the weighting function plotted in Fig.~\ref{Fig:weighting} for the
two cases is much larger in the muon case. 
The lepton mass acts, in fact,  as a cutoff scale.
Nonetheless, the B$\chi$PT result obtained hereby demonstrates that the magnitude of this effect is
not nearly enough to explain the `proton radius puzzle', which amounts to a discrepancy of about 300 $\mu$eV.

As seen from~Table~\ref{Table:Summary}, our B$\chi$PT result for the polarizability effect agrees with the previous evaluations
based on dispersive sum rules, but is substantially smaller in magnitude 
than the HB$\chi$PT result of Nevado and Pineda~\cite{Nevado:2007dd}.
This is of course not the first case when the B$\chi$PT and HB$\chi$PT results differ significantly --- the polarizabilities themselves provide such an example. 

The differences between HB$\chi$PT and  B$\chi$PT results are often interpreted as the uncertainty of $\chi$PT calculations. This is interpretation
is too naive as there are physical effects that distinguish the two. 
For example, the 
B$\chi$PT calculations obey analyticity exactly while the HB$\chi$PT ones only approximately. Furthermore, we have checked that in HB$\chi$PT the contribution
from momenta beyond the $\chi$PT applicability domain is 
somewhat bigger than the expected uncertainty of the calculation. The B$\chi$PT result is more ``effective" in this respect, as the high-momentum contribution therein is well
within the expected uncertainty.

Within the B$\chi$PT calculation, we have verified the dispersive sum rules given
in \Eqref{dispSRs} and confirmed the statement of Ref.~\cite{Birse:2012eb}   
that the split between the `elastic' and `inelastic' 2$\ga$ 
contributions corresponds unambiguously to the split between the Born and non-Born
parts of the VVCS amplitude, rather than between the pole and non-pole parts.

We have observed that the $\De(1232)$-excitation mechanism shown in  \Figref{Delta} does not impact the Lamb shift in 
a significant way because the dominant magnetic-dipole ($M1$) transition is suppressed, as is the entire 
magnetic polarizability effect. The $\De(1232)$-excitation effect is however important for the dispersive
calculation because it is prominent in the proton structure functions and hence must be included in the
`subtraction' contribution to achieve a consistent cancelation of the $M1$ $\De(1232)$ excitation. In 
most of the models this is roughly achieved by using an empirical value for the magnetic polarizability which
includes the large paramagnetic effect of the $M1$ $\De(1232)$ excitation. In the HB$\chi$PT-inspired calculation of 
the `subtraction' term \cite{Birse:2012eb} the $\De$-excitation is not included, however the situation is ameliorated 
by the low-energy constant from $O(p^4)$, which is chosen 
to reproduce the empirical value of the magnetic polarizability.

Naive dimensional analysis shows that $\chi$PT at leading order is capable of yielding predictions for the entire
two-photon correction to the Lamb shift. The polarizability part of that correction has been considered in this work. The last row of the last column of 
Table~\ref{Table:Summary} contains the $O(p^3)$ B$\chi$PT
prediction for the proton polarizability effect on the $2S$-level of 
$\mu$H. One needs to add to it the `elastic' contribution (or, alternatively, the $3^\mathrm{rd}$ Zemach moment together with `finite-size recoil'), to obtain the full 
$O(\al_{em}^5)$ effect of the proton structure in $\mu$H Lamb shift.
Using an empirical value for the `elastic' contribution from
Ref.~\cite{Birse:2012eb} [i.e., $-24.7(1.6)$ $\mu$eV], our result for the full
$2\ga$ contribution to the $2P\,$--$\, 2S$ Lamb shift is in nearly prefect agreement with the presently favored value 
\cite{Antognini:2013rsa,Birse:2012eb} of $33(2)$ $\mu$eV.

While the leading-order $\chi$PT calculation gives a reliable prediction for the
polarizability contribution, the splitting of it into `inelastic' and `subtraction' works less well, because of 
the missing $\Delta$(1232)-excitation effect, which will only enter at the (future) next-to-leading order calculation.
Indeed, $\chi$PT is capable of providing results for the Lamb
shift contribution beyond $O(p^3)$. The main difficulty then is to
include all the appropriate operators from the effective lepton-nucleon Lagrangian, with corresponding low-energy constants
fixed to the two-photon exchange component of the low-energy lepton-nucleon scattering. It will therefore be interesting but very difficult to carry
out any beyond-the-leading-order calculation in a systematic way.
 

\section*{Acknowledgements}

It is a pleasure to thank M.~Birse, C.~E.~Carlson, M.~Gorchtein, R.~J.~Hill,  S. Karshenboim, N.~Kivel, J.~McGovern, G.~A.~Miller,
A.\ Pineda, M.\ Vanderhaeghen, and T.\ Walcher for insightful, often inspiring,  discussions and communications. We furthermore thank A.~Antognini, M.~Birse, C.~E.~Carlson, M.~Gorchtein, J.~McGovern, R.\ Pohl, M.\ Vanderhaeghen for helpful remarks
on the manuscript.
This work was partially supported by the Deutsche Forschungsgemeinschaft (DFG) through the Collaborative Research Center
``The Low-Energy Frontier of the Standard Model" (SFB 1044), by
 the Cluster of Excellence ``Precision Physics, Fundamental Interactions and Structure of Matter" (PRISMA),
and by the UK Science and Technology Facilities Council through the grant ST/J000159/1.
V.~L.\ thanks the Institut f{\"u}r Kernphysik at the Johannes-Gutenberg-Universit{\"a}t Mainz for their kind hospitality.

\appendix

\section{Non-Born amplitudes of zero-energy VVCS} \label{App:VVCS-amplitudes}

Here we specify the  VVCS amplitudes at $\nu=0$. 
The expressions are given in terms of dimensionless variables:
the pion-proton mass ratio $\mu=m_{\pi}/M_p$ and the momentum-transfer $Q$ expressed in the proton mass units.
The pre-factor contains the fine-structure constant $\alpha_{em}\simeq 1/137.036$, 
the proton  mass $M_p\simeq 938.3$ MeV, the nucleon axial coupling $g_A\simeq 1.27$ and the pion decay constant $f_\pi\simeq 92.4$~MeV. We neglect the isospin breaking effects, such as differences in the nucleon or pion masses. For the latter we assume $m_\pi\simeq 139$ MeV.

The $O(p^3)$  
B$\chi$PT expressions are given by:
{\small
\begin{align}
T_1^{(\mathrm{NB})}(0,Q^2)&=-\frac{\alpha_{em} g_A^2 M_p}{2 \pi f_{\pi }^2} \int_{0}^{1}\!\!dx \int_{0}^{1}\!\!dy\ \left\{\sqrt{\frac{4 \mu ^2}{Q^2}+1} \log \left(\frac{\sqrt{(4 \mu ^2/Q^2)+1}+1}{\sqrt{(4 \mu ^2/Q^2)+1}-1}\right)\right. \nonumber \\
                 &+\frac{3 (x-1)}{Q^2} \left[\log \left(Q^2 (-(x-1)) x+\mu ^2 x+(x-1)^2\right)-\log\left(x^2+\left(\mu ^2-2\right) x+1\right)\right]  \nonumber \\
                 &- \frac{2 (x-1)^2 x \left[(x-1)^2 \left(Q^2 y^2-1\right)-\mu ^2 x\right]}{\left[(x-1)^2 \left(Q^2 (y-1)    y-1\right)-\mu ^2 x\right] \left[(x-1) \left(Q^2 (x-1) y^2+Q^2 y-x+1\right)-\mu ^2 x\right]} \nonumber \\
                 &+\frac{(x-1)^2 (y-1) \left[(x-1) \left(Q^2 (x-1) y^2-Q^2 (x-2) y+x-1\right)-\mu ^2
   x^2\right]}{\left[(x-1) \left(Q^2 (x-1) y^2+Q^2 y-x+1\right)-\mu ^2 x\right]^2} \nonumber\\
               & -\frac{4 x^2 (x-1) (y-1)}{x^2 \left(Q^2 y^2-1\right)-x \left(\mu ^2+Q^2
   y-2\right)-1}-\frac{4 x (x-1)^2}{x^2 \left[Q^2 (y-1) y-1\right] -\left(\mu ^2-2\right) x-1} \nonumber \\
               &\left.+\frac{2 x (x-1)}{x^2+\left(\mu ^2-2\right) x+1}-2\right\}\,,
\end{align}
\begin{align}
T_2^{(\mathrm{NB})}(0,Q^2)&=-\frac{\al_{em} g_A^2 M_p}{\pi  f_{\pi }^2} \int_{0}^{1}\!\!dx \int_{0}^{1}\!\!dy\  \biggl\{\frac{(x-1)^2 x (y-1) \left[(x-1) \left(-Q^2 y+2 x-2\right)+\mu ^2 x\right]}{\left[(x-1) \left(Q^2 (x-1) y^2+Q^2 y-x+1\right)-\mu ^2 x\right]^2} \nonumber \\
                  & + \frac{4 (x-1) x^2 y \left[x^2 \left(Q^2 (y-1) y+1\right)-\left(\mu ^2+2\right) x+1\right]}{\left[x
   \left(-\mu ^2+x \left(Q^2 (y-1) y-1\right)+2\right)-1\right] \left[x^2 \left(Q^2 y^2-1\right)-x
   \left(\mu ^2+Q^2 y-2\right)-1\right]} \nonumber \\
                 & +\frac{4 x}{Q^2} \left[\log \left(Q^2 x y (1-x y)+\mu ^2 x+(x-1)^2\right)\right.\nonumber \\
                 & \hphantom{+\frac{4 x}{Q^2}[}\left.-\log \left(x \left(\mu ^2+x \left(1-Q^2 (y-1) y\right)-2\right)+1\right)\right] \nonumber \\
               & +\frac{4 (x-1) x^3 (y-1) \left[Q^2 y (x y-1)-\mu ^2\right]}{\left[x \left(\mu ^2+Q^2 (-x) y^2+Q^2 y+x-2\right)+1\right]^2} \nonumber \\
               & +\frac{2 (x-1)}{Q^2} \biggl[\frac{(x - 1)^2 (Q^2 (y - 1) y + 1)}{\left[(x-1)^2 \left(Q^2 (y-1) y-1\right)-\mu ^2 x\right]} \nonumber \\
               & \hphantom{+\frac{2 (x-1)}{Q^2} \biggl[}  -\frac{(x - 1)^2 (Q^2 (y - 1) y + 1)}{\left[(x-1) \left(Q^2 (x-1) y^2+Q^2 y-x+1\right)-\mu ^2 x\right]}  \nonumber \\
              &  \hphantom{+\frac{2 (x-1)}{Q^2} \biggl[} -\log \left[Q^2 (1-x) y ((x-1) y+1)+\mu ^2 x+(x-1)^2\right] \nonumber \\
              & \hphantom{+\frac{2 (x-1)}{Q^2} \biggl[} +\log \left[\mu ^2 x-(x-1)^2 \left(Q^2 (y-1) y-1\right)\right]\biggr] \nonumber \\
              &   -\frac{3}{Q^4} \biggl[-\frac{2 Q^2 x (x-1)^2}{x \left(\mu ^2+x-2\right)+1}-\left[\left(Q^2-2\right) x+2\right] \log \left[x \left(\mu
   ^2+x-2\right)+1\right] \nonumber \\
             &  \hphantom{-\frac{3}{Q^4} \biggl[}+\left[\left(Q^2-2\right) x+2\right] \log \left[x \left(\mu ^2+Q^2 (1-x)+x-2\right)+1\right]\biggr]       \biggr\}\,.
\end{align}
}

\section{Tree-level electroproduction cross sections}\label{App:CrossSections}

\begin{figure*}
\begin{center}
\epsfig{file=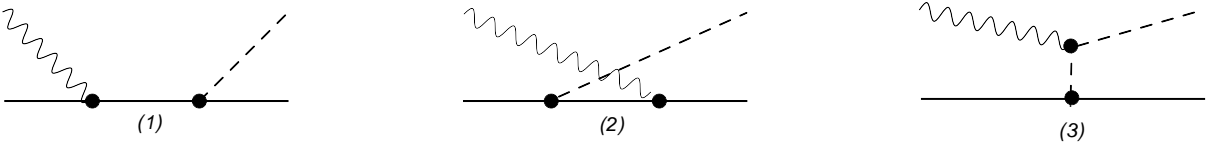,width=16cm,angle=0} 
\caption{Graphs for pion electroproduction amplitude at leading order.
The $\pi NN$ couplings are pseudo-scalar as derived from the transformed Lagrangian Eq.~(\ref{Eq:expNlagran2}).
\label{Fig:DiagsOp}}
\end{center}
\end{figure*}

Here we present our results for the electroproduction cross sections
corresponding to  diagrams in Fig.~\ref{Fig:DiagsOp}.
We give them in terms of the following
dimensionless variables:

{\small 
\begin{align}
&\alpha_\gamma= (E_i^{N})_\mathrm{cm}/\sqrt{s}=\frac{s+M_p^2+Q^2}{2 s}\,,   \\
&\alpha_\pi= (E_f^{N})_\mathrm{cm}/\sqrt{s} = \frac{s+M_p^2-m_\pi^2}{2 s}\,,  \\              
 & \beta_\gamma = E^{\gamma}_\mathrm{cm}/\sqrt{s} = \frac{s-M_p^2-Q^2}{2 s}\,,   \\
 & \beta_\pi= E^{\pi}_\mathrm{cm}/\sqrt{s} = \frac{s-M_p^2+m_\pi^2}{2 s}\,,  \\
 &\lambda_\gamma  = |\vec{q}_i|_\mathrm{cm}/\sqrt{s}= \frac{\sqrt{(s-M_p^2 - Q^2)^2+4 s Q^2}}{2 s}\,,  \\
&\lambda_\pi =  |\vec{q}_f|_\mathrm{cm}/\sqrt{s}  = \frac{\sqrt{(s-M_p^2 + m_\pi^2)^2-4 s m_\pi^2}}{2 s}\,, 
\end{align} 
}
where $(E_i^{N})_\mathrm{cm}$ is the energy of the incoming nucleon, $(E_f^{N})_\mathrm{cm}$ is the energy of the outgoing nucleon, $E^{\gamma}_\mathrm{cm}$ the energy of the incoming photon, $E^{\pi}_\mathrm{cm}$ the energy of the outgoing pion, $|\vec{q}_i|_\mathrm{cm}$ the relative three-momentum of the incoming particles and $|\vec{q}_f|_\mathrm{cm}$ the relative three-momentum of the outgoing particles, all in the centre-of-mass frame (CM).

We show below the results obtained for the pion electroproduction cross sections for the different channels. 
They have been calculated by using the energy of the incoming virtual photons in the laboratory frame as the flux factor of the incoming particles.
We have checked that they reproduce the result at the real photon point shown in Refs.~\cite{Lensky:2009uv}.
As in Appendix \ref{App:VVCS-amplitudes}, $Q$ and $s$  are
in the units of proton mass.


{\footnotesize

\begin{align}
&\sigma_T^{(\pi^+ n)}=\frac{\alpha_{em} g_A^2 \lambda_\pi }{4 f^2_\pi s^2 (s-1+Q^2)\lambda_\gamma^3} \left\{ \frac{2 s \lambda_\gamma}{(s-1)^2}\left[ 2 \mu^2((s-1)^2-Q^2 s \lambda_\gamma^2) + (1-s)(Q^4+2Q^2 s \beta_\gamma \beta_\pi \nonumber \right. \right.\\
&\left.\left. + 2 s (1 -s+2 s \beta_\gamma \beta_\pi) \lambda_\gamma^2)\right]  +\frac{1}{(s-1)\lambda_\pi} \left[  2 \mu^2 (1-s) (Q^2+2 s \beta_\gamma \beta_\pi)+Q^2((Q^2+2 s\beta_\gamma \beta_\pi)^2 \right. \right. \nonumber \\
 & \left.\left.-4 s^2 \lambda_\pi^2 \lambda_\gamma^2) \right] \arctanh\left[ \frac{2 s \lambda_\pi \lambda_\gamma}{Q^2 + 2 s \beta_\gamma \beta_\pi} \right]\right\}\,,
\end{align}

\begin{align}
&\sigma_T^{(\pi^0 p)}=\frac{\alpha_{em} g_A^2\lambda_\pi}{2 f_\pi^2 (s-1+Q^2) (s-1)^2 } \left\{ \frac{1}{-2 s(1 + s (-1 + 2 \beta_\gamma \beta_\pi))^2 \lambda_\gamma^2 + 8 s^3 \lambda_\pi^2 \lambda_\gamma^4}\left[ (1 - s)(Q^2 (s-1  \right.\right. \nonumber\\
& -2 s \beta_\gamma \beta_\pi ) - 2 s (s-1 + 2 s \beta_\gamma \beta_\pi) \lambda_\gamma^2) ((1+s (-1 + 2 \beta_\gamma \beta_\pi))^2 - 4 s^2 \lambda_\pi^2 \lambda_\gamma^2) + 2 \mu^2 (-(s-1)^2 (1 \nonumber\\
&\left. + s(-1 + 2 \beta_\gamma \beta_\pi))^2 + 2 s (Q^2 (1 + 2 s(-1 + \beta_\gamma \beta_\pi) + s^2 (1 + 2 \beta_\gamma \beta_\pi (-1 + \beta_\gamma \beta_\pi))) + 2 (s-1)^2 s \lambda_\pi^2 )\lambda_\gamma^2 \right.  \nonumber \\
&\left. -4 Q^2 s^3 \lambda_\pi^2 \lambda_\gamma^4 ) \right]+ \frac{1}{4 s^2 \lambda_\pi \lambda_\gamma^3}(1-s) \left[-((2 \mu^2 +Q^2)(1 - s) + 2 Q^2 s \beta_\gamma \beta_\pi)(1+s(-1+2 \beta_\gamma \beta_\pi))\right.\nonumber \\
&\left.\left.+2s (1 - 2 s +s^2 + 2 Q^2 (\mu^2 + s \lambda_\pi^2))\lambda_\gamma^2\right]\arctanh\left[ \frac{2 s \lambda_\pi \lambda_\gamma}{1 + s (-1 + 2 \beta_\gamma \beta_\pi)}\right]\right\}\,,
\end{align}

\begin{align}
&\sigma_{L}^{(\pi^+ n)}=\frac{\alpha_{em} g_A^2 \lambda_\pi}{2 f_\pi^2 Q^2 \, (s-1+Q^2) (s-1)^2 \lambda_\gamma^3}\left\{   \frac{1}{(Q^2 + 2 s \beta_\gamma \beta_\pi )^2 - 4 s^2 \lambda_\pi^2 \lambda_\gamma^2}    \left[ 2 \lambda_\gamma (-Q^2(1-s)(\beta_\gamma^2 (Q^2  \right.\right.\nonumber\\
& +2 s \beta_\gamma \beta_\pi) + (1 + s (-1 + 2 \beta_\gamma(-1 + 2 \alpha_\pi+ \beta_\pi)))   \lambda_\gamma^2)   ((Q^2+2 s \beta_\gamma \beta_\pi)^2 - 4 s^2 \lambda_\pi^2 \lambda_\gamma^2) + \mu^2 (-2 (s-1)^2  \nonumber\\
& \times \beta_\gamma^2 (Q^2+2s \beta_\gamma \beta_\pi)^2+ (Q^8 + 4 Q^6 s \beta_\gamma \beta_\pi - 4 Q^2 (1 - s) s \beta_\gamma ((1-s)(-1+\alpha_\pi)+4 s \beta_\gamma \beta_\pi)  \nonumber \\
&+ 4 Q^4 s \beta_\gamma (s-1 + s \beta_\gamma \beta_\pi^2) +4 s^2 (s-1) \beta_\gamma^2  (2 \beta_\pi ((1 - s) (-1+\alpha_\pi) + 2 s \beta_\gamma \beta_\pi) + (s-1) \lambda_\pi^2) )\lambda_\gamma^2  \nonumber \\
&\left. - 4 s^2 ((1 +s^2 )(\alpha_\pi-1)^2 + (Q^2 + 2 s \beta_\gamma \beta_\pi)^2+ (Q^4 + 4 s^2 \beta_\gamma) \lambda_\pi^2 - 2 s ((\alpha_\pi-1)^2 + 2 \beta_\gamma \lambda_\pi^2) )\lambda_\gamma^4 \right. \nonumber \\
&\left. + 16 s^4 \lambda_\pi^2 \lambda_\gamma^6   )   ) \right]   + \frac{1 - s}{s\lambda_\pi}  \left[   \beta_\gamma (Q^2 + 2 s \beta_\gamma \beta_\pi) + 2 s (\alpha_\pi-1) \lambda_\pi^2    \right] \left[  \beta_\gamma (Q^4 + 2 \mu^2 (1 - s) + 2 Q^2 s \beta_\gamma \beta_\pi  )\right. \nonumber \\
&\left.\left. + 2 s (2 \mu^2 + Q^2 \alpha_\pi) \lambda_\gamma^2     \right]      \arctanh\left[ \frac{2 s \lambda_\pi \lambda_\gamma }{Q^2+ 2 s \beta_\gamma \beta_\pi}\right]  \right\}\,,
 \end{align}

\begin{align}
&\sigma_{L}^{(\pi^0 p)}=\frac{\alpha_{em} g_A^2 \lambda_\pi}{4 f_\pi^2 Q^2\, (s-1+Q^2)(s-1)^2 \lambda_\gamma^3}\left\{  \frac{1}{(1+s(-1+2 \beta_\gamma \beta_\pi))^2 -4 s^2 \lambda_\pi^2 \lambda_\gamma^2}\left[  4\mu^2 \lambda_\gamma (-(1 -s )^2  \right.\right.\nonumber \\
&\times\beta_\gamma^2 (1 + s (-1+2\beta_\gamma \beta_\pi))^2+ (-2 s (1+\alpha_\pi) \beta_\gamma + (Q^4 + 2 s^2 \beta_\gamma (3+3\alpha_\pi - 4 \beta_\gamma \beta_\pi - 2 \alpha_\pi \beta_\gamma \beta_\pi + \beta_\gamma \lambda_\pi^2)) \nonumber \\
&+ s^2 (Q^4 (1+2 \beta_\gamma \beta_\pi (-1 + \beta_\gamma \beta_\pi)) + 2 s^2 \beta_\gamma (\alpha_\pi - 2\alpha_\pi \beta_\gamma \beta_\pi + (1-2 \beta_\gamma \beta_\pi)^2 + \beta_\gamma \lambda_\pi^2) ) - 2 s (Q^4 (1-\beta_\gamma \beta_\pi)  \nonumber \\
&+s^2 \beta_\gamma (3+\alpha_\pi (3-4\beta_\gamma \beta_\pi)+2 \beta_\gamma (2 \beta_\pi (\beta_\gamma \beta_\pi -2) + \lambda_\pi^2)))) \lambda_\gamma^2 - 2 s^2 ((1+\alpha_\pi^2) + Q^4 \lambda_\pi^2 - 2 s (1+\alpha_\pi^2  \nonumber\\
&- 2 \beta_\gamma \beta_\pi + 2 \beta_\gamma \lambda_\pi^2)+ s^2 (\alpha_\pi^2+(1-2 \beta_\gamma \beta_\pi)^2 + 4 \beta_\gamma \lambda_\pi^2)) \lambda_\gamma^4 + 8 s^4 \lambda_\pi^2 \lambda_\gamma^6)  - 2 Q^2 (1 - s) \lambda_\gamma (s \beta_\gamma^2 (-1+2 \beta_\gamma \beta_\pi)  \nonumber \\
&\left.+ s (1+2\beta_\gamma (-2 + 2 \alpha_\pi +\beta_\pi)) \lambda_\gamma^2 +\beta_\gamma^2-\lambda_\gamma^2) (1 +s (-1+ 2 \beta_\gamma \beta_\pi - 2 \lambda_\pi \lambda_\gamma )) (1 + s(-1 +2 \beta_\gamma \beta_\pi+2 \lambda_\pi \lambda_\gamma))\right]  \nonumber \\
&+\frac{1-s}{s \lambda_\pi} \left[   2\mu^2 (-Q^4 \lambda_\gamma^2+((1-s) \beta_\gamma + 2 s \lambda_\gamma^2)( \beta_\gamma+ s \beta_\gamma(-1+2 \beta_\gamma \beta_\pi) +2 s \alpha_\pi \lambda_\gamma^2)) + Q^2 ((\beta_\gamma-\lambda_\gamma) \right. \nonumber \\
&\left.+ s (\beta_\gamma (-1+2 \beta_\gamma \beta_\pi) + \lambda_\gamma+2(\alpha_\pi -1)\lambda_\gamma^2  )) ( (\beta_\gamma+\lambda_\gamma) + s (\beta_\gamma (-1+2 \beta_\gamma \beta_\pi) + \lambda_\gamma(-1+2 (\alpha_\pi-1)\lambda_\gamma) )  ) \right] \nonumber \\
&\times\left.\arctanh\left[  \frac{2 s \lambda_\pi \lambda_\gamma}{1+ s (-1+2 \beta_\gamma \beta_\pi)} \right]\right\}\,.
\end{align}
}

\end{document}